\documentclass[paper]{aaprep}

\usepackage{amsfonts}
\usepackage{graphicx}
\usepackage{natbib}
\bibpunct{(}{)}{;}{a}{,}{,}

\def\e{\delta t}
\def\a{\alpha_{s}}

\def\muas{\mu as}

\def\OO#1{{\cal O}(#1)}
\newcommand{\vecg}[1]{\mbox{\boldmath$#1$}}
\newcommand{\ve}[1]{\vecg{#1}}

\def\s{\ve{\sigma}}
\def\xlrm{\ve{x}_{LRM}^0}
\def\Xlrm{x_{LRM}}
\def\lO{\ve{l_{0}}}

\def\vlrm{\ve{v}_{LRM}^0}
\def\V{\ve{V}_{orb}}
\def\xobs{\ve{x}_{obs}}
\def\xe{\ve{x}_{e}}
\def\D{\ve{D}}
\def\DD{\ve{\Delta D}}

\def\DDte{\D\left[\te\right] - \D\left[\teO\right]}

\def\DDteto{\Delta\D\left[\tobs\right]}

\def\Dxobs{\ve{\Delta x}_{obs}\left[\tobs\right]}
\def\xs{\ve{x_{s}}}
\def\pivec{\ve{\pi}}
\def\xe{\ve{x_{e}}}

\def\rvec{\ve{r}}
\def\X{\times}

\def\ap{\simeq}

\def\tobs{t_{obs}}
\def\tobsO{t^{0}_{obs}}
\def\te{t_{e}}
\def\teO{t^{0}_{e}}
\def\Dtobs{\Delta t_{obs}}
\def\Dte{\Delta t_{e}}
\def\Dt{\Delta t}
\def\xobsO{\ve{x}_{obs}^0}

\def\pvec{\ve{p}}
\def\qvec{\ve{q}}

\begin{document}

\title{Astrometric Light-Travel Time signature of sources
in nonlinear motion} \subtitle{ I.Derivation of the effect and
radial motion}
\author{Guillem Anglada \and Jordi Torra}

\institute{ Departament d'Astronomia i Meteorologia, Universitat de
Barcelona, Av. Diagonal 647, 08028 Barcelona, Spain}

\date{Received \today / Accepted \today}

\offprints{Guillem Anglada, \email{anglada@am.ub.es}}

\abstract{Very precise planned space astrometric missions and recent
improvements on imaging capabilities require a detailed review of
the assumptions of classical astrometric modeling.}{We show that
Light-Travel Time must be taken into account to model the kinematics
of astronomical objects in nonlinear motion, even at stellar
distances.}{A closed expression to include Light-Travel Time in the
actual astrometric models with nonlinear motion is provided. Using a
perturbative approach the expression of the Light-Travel Time signature
is derived. We propose a practical form of the
astrometric modelling to be applied in astrometric data reduction
of sources at stellar distances($d>1 pc$).}{We show that the
Light-Travel Time signature is relevant at $\muas$ accuracy (or even
at $mas$) depending on the time span of the astrometric measurements.
We explain how information about the radial motion of a source can be 
obtained. Some estimative numbers are provided for
known nearby binary systems}{In the light of the obtained results,
it is clear that this effect must be taken into account to interpret
any kind of precise astrometric measurements. The effect is particularly
interesting in measurements performed by the planned astrometric space missions
(GAIA, SIM, JASMINE, TPF/DARWIN). Finally an objective criterion
is provided to quickly evaluate whether the Light-Travel Time modeling is
required for a given source or system.}

\keywords{Astrometry -- Stars: kinematics -- Reference systems --
Time -- Stars: binaries -- Stars: planetary systems}

\titlerunning{Astrometric LTT signature and nonlinear motion}
\authorrunning{G.~Anglada, J.Torra}

\maketitle

%\begin{document}

\section{Introduction and notation} \label{sec:intro}

The new capabilities for producing very precise astrometric
measurements come essentially from space-borne astrometric
missions like SIM\citep{SIM:1998}, GAIA\citep{Perryman:et:al:2001}
or JASMINE \citep{JAZMINE:2002}, and require a revision of the classic
astrometric assumptions at any level. Some of the concepts that are
being reviewed carefully are those involving light signal
propagation \citep{klioner:basic,leponcin:2004} and the description
of the astronomical sources and observers \citep{klioner:2004} in
the context of the IAU resolutions \citep{brumberg:BCRS} that aim
to define a consistent framework to model the astronomical
observations.

This effort involves many groups and individuals all around
the world and one relevant aspect is the appropriate description of 
stellar motion. This work focuses on the impact of
Light-Travel Time (LTT) on the observed direction of a source
outside the solar system. Such considerations are as old
as modern astronomy itself; in the 17$^{th}$ century Ole R\"{o}mer used it 
to give the first estimation of the velocity of light.

In solar system dynamics, light travel delays are already widely
considered and applied. Our aim is to show that it is also relevant for
distant objects($d > 1\ pc$) and to develop analytic expressions to
include the astrometric LTT signature in the ultraprecise
astrometric modeling. We will provide an expression which is
algorithmically efficient and fully compatible with the IAU
standards \citep{brumberg:BCRS} to the required level of accuracy. A
full scheme of a more general astrometric model compatible with the
BCRS is given in \citet{klioner:basic}. The present work may be seen
as a refinement of the formulas given there to generalize stellar
motion.

The first two Sections are devoted to establishing the physical framework 
and defining the relevant quantities. Section \ref{sec:delay} is devoted to
determining the relation between the emission time interval
$\Dte$ and the observation time interval $\Dtobs$ to the required
accuracy for astrometric purposes.

The baseline astrometric model is stated in Section \ref{sec:LRM}
where the expression for a point source in linear motion is developed.
There we define the concept of \textit{Linear Reference
Motion}(LRM) which will be very useful in the later developments.
The same development for the source in linear motions was proposed
by \citet{kopeikin:1992}, where the LTT due to observer's position
was already included. Light-Travel Time effects on sources in linear
motion and their relation to constant radial velocities is a topic
extensively discussed in the bibliography and directly related to
apparent superluminal motion. A good review of this issue is found
in \citet{lindegren:2003}. In Section \ref{sec:NLM}, the astrometric
model is naturally extended to sources in nonlinear motion.

Section \ref{sec:analytic} is devoted to obtaining a quantitative
description of the astrometric LTT signature for a point source. The LTT
signature is found to appear as a second order correction (as, for
example, the \emph{astrometric radial velocity} described in
\citet{lindegren:2003}).

In Section \ref{sec:real}, a series of examples shows how the LTT
signature carries information about the radial geometry of the
trajectory of a source. This is of particular interest in binary
systems or any kind of objects in Keplerian orbits(such as
exoplanetary systems), because resolved LTT signatures may lead to
the determination of the full set of orbital parameters without
spectroscopic measurements, as is commonly required
\citep[see][chap.~1]{batten:1973}.

Section \ref{sec:conclusions} provides an heuristic relation
to evaluate the significance of the LTT effects in the geometrical
characterization of any kind of astronomical structures.

The use of LTT effects in epoch observations to determine properties
of sources at stellar distances was first proposed by
\citet{Irwin:1952}, whose work is applicable to binary systems with
at least one variable component. The reader is encouraged to check
\citet{ribas:LTT} as an example of the use of this technique.
Similar and more sophisticated models are used in precise pulsar
timing. One of the most spectacular results using accurate LTT
modeling is the detection of the first exoplanetary system
around another star (a pulsar indeed) by \cite{pulsar:1992}.

Let us summarize the most important notation and conventions used
in the present the paper:

\begin{itemize}

\item The velocity of light is c.

\item  The three-dimensional coordinate quantities (" 3-vectors ")
referring to the spatial axes of the corresponding reference system
are set in boldface italic: $\ve{a} = a^i \hat{\ve{e}}_i$, where $\{\,
\hat{\ve{e}}_i\, ;\, i=1,2,3 \}$ is the spatial basis defined
by the coordinate system of the BCRS.

\item  The scalar product of two vectors $\ve{a}$ and $\ve{b}$
is denoted by $\ve{a} \cdot \ve{b} = \sum_{i=1}^{3}a^i b^i$

\item  The \emph{Euclidean norm} of a 3-vector $\ve{a}$
is denoted $\| \ve{a}\|$ and is computed as $\| \ve{a}\| =
\left(\ve{a} \cdot \ve{a}\right)^{1/2}$.

\item  $\left< \ve{a} \right>$ means that the vector
must be normalized using its own Euclidean norm $\left< \ve{a}
\right> = \frac{\ve{a}}{\| a\|}$.

\item The components of the obtained vector using the cross product operation $\ve{a} \times
\ve{b}$ are given by $\left(\ve{a}\times \ve{b}\right)^i =
\varepsilon_{ijk}a^{j}b^{k}$, where $\varepsilon_{ijk}$ is the
completely antisymmetric \emph{Levi-Civita} symbol with
$\varepsilon_{xyz}=+1$.

\item All physical quantities are expressed using the SI units (or MKS) if no
particular comment is added in the text.

\item Small angles are usually given in fractions of \emph{arcseconds}. The most
common shortcuts are $mas = 10^{-3}\,\arcsec$ and $\mu as = 10^{-6}\, \arcsec$.

\item A symbol $p$ inside square brackets $\left[\ldots\right]$ after a
symbol $f$, means that $f[p]$ is an explicit function of $p$. This
notation is used thoughout the paper where the arguments in a function
may appear ambiguous to the reader. Round brackets
$\left(\ldots\right)$ are exclusively used to group algebraic expressions.

\end{itemize}

\section{Trajectories, quantities and reference system} \label{sec:trajectories}

Our purpose is to determine the observed direction of a moving object
from the position of an observer at rest with respect to the
barycenter of the solar system in absence of gravitational fields.
To do that one must describe the motion of a source in
terms of the observation instant $\tobs$ instead of the emission
instant $\te$. The relation between an emission time interval $\Dte$
and its corresponding observation time interval $\Dtobs$ will be
nonlinear and time dependent due to the nonlinear change of the
distance a light signal must cover. This nonlinear time dependence
will add additional apparent nonlinear terms to its motion on the
celestial sphere.

We restrict the discussion to a particular inertial frame of special 
relativity where the space-time metric
is assumed to be \emph{Minkowsky} metric with signature $\left( \ -
\ + \ + \ + \right)$. All the quantities and vectors refer to
the BCRS spatial coordinates $x^i$ and the time coordinate $t$
describing the events is TCB -- see \citet{brumberg:BCRS}. The BCRS
metric is not a \emph{Minkowsky} metric, but since the astrometric
LTT signature is already very small the gravitational light bending
and the kinematical aberration can be treated as \emph{a posteriori}
effects to the observed direction (see \citet{klioner:basic}). The
trajectory of a point source is described as a Linear Reference
Motion $\ve{x}_{LRM}\left[t\right]$ plus a nonlinear shift
$\D\left[t\right]$. The spatial part of the trajectory of the
\textit{source} in BCRS coordinates is given by

\begin{eqnarray} \label{eq:track}
\xs\left[t\right] &=&\ve{x}_{LRM}\left[t\right] + \D\left[t\right]\ , \\
\ve{x}_{LRM}\left[t\right] &=& \xlrm + \vlrm \left(t - \teO \right)\ ,
\end{eqnarray}

\noindent where $\xlrm$ are the coordinates of the LRM at some
instant $\teO$. This initial instant will be discussed more
precisely below. Formally, the constant velocity term $\vlrm$ could
be included in $\D\left[t\right]$, but it is very useful to keep it
apart in order to define properly the \textit{Barycentric astrometric
parameters}(see Section \ref{sec:LRM}) and relate them to the
physical quantities in (\ref{eq:track}). As an example, $\xlrm$ and 
$\vlrm$ describe the motion of the center of mass 
of a binary system and $\D$ describes the orbital motion of one of 
the components.

The value of coordinate time $t$ at the emission event $E$ is
denoted by $\te = t\left[E\right]$. The spatial coordinates $\xs$ at
the emission event (which coincide with the spatial coordinates of
the \textit{source}) are denoted by $\xs\left[\te\right]$. In the
same way the value of the coordinate time at the observation event
is $t\left[Obs\right]=\tobs$ and the spatial coordinates of such
an event are $\xobs\left[\tobs\right]$. Please note that
$\te$ and $\tobs$ are both given in the same time scale, which is TCB. 

Since in Minkowsky space-time the light rays follow straight lines,
the spatial vector joining an event of emission at $\te$ and an
event of observation at $\tobs$ defines the \textit{observed unit
direction} as

\begin{eqnarray}
\rvec &=& \left< \xs\left[\te\right] -
\xobs\left[\tobs\right]\right>.
\end{eqnarray}

\begin{figure}
\centering
\includegraphics[width=6.0cm]{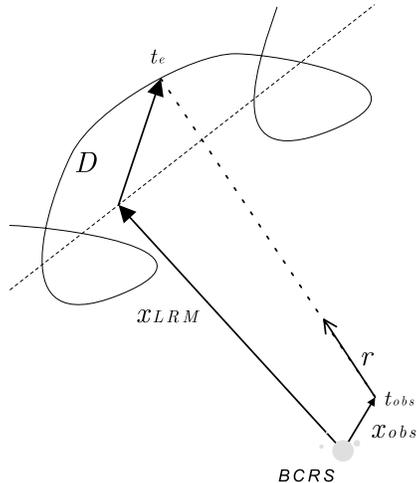}
\caption{Scheme of the vectors involved in the computation of the
observed direction at a given event of observation. The vector
$\ve{x}_{LRM}$ is used to define a fiducial Linear Reference Motion,
which could be, for example, the trajectory of the center of mass of
a binary system. The shift vector $\D$ is the nonlinear contribution
to the motion. Using again the example of the binary systems, it
might be the orbital motion of a component around the center of mass
of the system. The little vector $\rvec$ is the observed direction
of the incoming light ray at the event of observation.}
\label{fig:scheme}
\end{figure}

\noindent For the later developments, we need to identify the
\emph{small time intervals}

\begin{eqnarray} \label{eq:smallintervals}
\OO1 \sim
\frac{\vlrm \Dt}{c} \ ; \
\frac{\D}{c}\ ;
\frac{\xobs}{c} \ . \
\end{eqnarray}

\noindent The vectors in (\ref{eq:smallintervals}) have dimensions
of time and have typical absolute values going from some minutes to
several hours or days (or even years). From now on, the module 
$\|\xlrm\|$ will be written as $\Xlrm$ to simplify the notation. It is 
also useful to introduce the \emph{small adimensional} quantities

\begin{eqnarray} \label{eq:small}
\OO1 \sim
\frac{\vlrm \Dt}{\Xlrm} \ ; \
\frac{\D}{\Xlrm}\ ;
\frac{\xobs}{\Xlrm} \ . \
\end{eqnarray}

\noindent Relations (\ref{eq:small}) impose
that the initial distance between the source and the observer
$\Xlrm$ must be a good deal larger than all the other time
dependent displacements $\vlrm \Dt$, $\D$ and $\xobs$. This
requirement is usual in the astrometric modelling of objects
beyond the solar system. At this point, it is useful to define the
unit vector

\begin{eqnarray} \label{eq:lO}
\lO = \frac{\xlrm}{\Xlrm}\ ,
\end{eqnarray}

\noindent which is the direction towards the position of the LRM
(i.e. the center of mass of a binary system) given by an observer at the
BCRS origin at a given reference instant $\tobsO$, usually called
\textit{Barycentric reference epoch}.

An important remark must made here. With the exception of the
initial direction parameterized through two angles, only the
parameters producing time dependent changes in the observed
direction will produce measurable effects. Despite the frequent 
appearance of $\teO$ throughout the paper, this parameter is not 
directly measurable. The physical quantities producing time 
dependent effects are $\Xlrm$, $\vlrm$, $\D$. They depend only 
on $\teO$ in their \emph{formal} definition implicitly given 
in (\ref{eq:track}).

\section{Equation of time delay} \label{sec:delay}

In the most general case, the interval of time between the event of
observation and the event of emission is related to the
\emph{spatial coordinate distance} as

\begin{eqnarray} \label{null}
\left(\tobs - \te \right) &=& \frac{1}{c}\|\xobs\left[\tobs\right] - \xs\left[\te\right] \| \\
&+& \Delta_{BCRS}\left[\te,\tobs\right] + \Delta_{ext}\left[\te,\tobs\right] \nonumber \\
&+& \OO2\ \nonumber.
\end{eqnarray}

\noindent The $\Delta$ terms on the right side can include
additional space-time effects due to gravitational contributions due
to the BCRS fields ($\Delta_{BCRS}$) and other external fields
$\Delta_{ext}$ that the photon may feel along its long trajectory.
This gravitational contribution is usually known as the Shapiro effect.
For a detailed discussion of the Shapiro effect see
\citet{kopeikin:1999}. We are interested in
the relation between the emission interval $\Dte$ and the
observation interval $\Dtobs$. Then, using the relation (\ref{null})
for two different events of emission $E^0$ and $E$ and their
respective observation events $Obs^0$ and $Obs$ we obtain

\begin{eqnarray} \label{eq:delay}
\Dtobs - \Dte &=& {\|\xobs\left[\tobs\right]  -\xs\left[\te\right]   \|\over{c}} \\
              &-& {\|\xobs\left[\tobsO\right] -\xs\left[\teO\right] \|\over{c}} \nonumber \\
              &+& \Delta_{BCRS}\left[ \te,\tobs \right] 
	        - \Delta_{BCRS}\left[\teO,\tobsO\right] \nonumber \\ 
	      &+& \Delta_{ext}\left[\te,\tobs\right] -
	          \Delta_{ext}\left[\teO,\tobsO\right] \nonumber \, ,
\end{eqnarray}

\begin{eqnarray}
\Dte   &=& \te - \teO \, , \\
\Dtobs &=& \tobs - \tobsO \, .
\end{eqnarray}

\noindent For most of the stars the absolute value of
$\Delta_{ext}$ may be very large since the gravitational fields of
the galaxies and other mass distributions may contribute to that;
however, in most circumstances, it will not change significantly during
the lifetime of a space astrometric mission (even in some hundreds
of years). An exception may be objects orbiting large concentrations
of mass or gravitational lensing events. In such cases, the model
for the observations must be carefully derived not only from the
point of view of the LTT. This might be the case for stars moving
close to the Milky way's central black hole (see \cite{ghez:1998}). An
example of such a detailed model is found in \cite{fragile:2000}.
Despite the use of such models to include very sophisticated space-time
effects, the astrometric LTT signature due to the nonlinear motion
is ignored. The contributions $\Delta_{BCRS}$ will heavily depend on the relative
position of the observer and the sources of gravitational fields
(i.e. Sun, planets). Considering that the physical diameter of such
bodies is some orders of magnitude larger that their Schwarschild
radius, the Shapiro term is going to add a very small time shift (a few
milliseconds as much). In such an interval of time, the direction of observation
will not change more than a few nanoarcseconds which is, by far, an
undetectable astrometric quantity with the current techniques.
From now on we will omit both $\Delta$ terms. 

The equation of time delay
(\ref{eq:delay}) depends on the module of the relative position of
the observer and the source, and in general, it cannot be used to
obtain a closed exact expression of $\Dte$ in terms of $\Dtobs$. A
perturbative approach is chosen here to obtain the more relevant
(first order) contributions, that is, relevant enough to affect the
observed direction at $\mu as$ level of accuracy.
With some algebra whose details are given in Appendix
\ref{sec:algebra1}, and using the definitions of small quantities
provided in (\ref{eq:smallintervals})--(\ref{eq:small}) it is
obtained that, at first order
\begin{eqnarray} \label{eq:simple}
\Dte = &\a& \left( \Dtobs \right. \\
                 &-&        \frac{1}{c}\, \lO \cdot \DDteto \nonumber \\
                 &+& \left.\frac{1}{c}\, \lO \cdot \Dxobs\  \right)\nonumber + \OO2 ;
\end{eqnarray}
\noindent In (\ref{eq:simple}) some notation shortcuts are applied. These are
\begin{eqnarray}
  \a &=& \frac{1}{1 + \lO \cdot {\vlrm \over{c}}}\, ,\label{eq:superlum}\\
  \Dxobs &=& \xobs\left[\tobs\right] - \xobs\left[\tobs^{0}\right], \label{eq:Dxobs}\\
  \DDteto &=& \D\left[\teO + \a\Dtobs\right] - \D\left[\teO\right]. \label{eq:DDteto}
\end{eqnarray}
\noindent The factor $\a$ multiplying the full
expression(\ref{eq:simple}) is the one responsible for apparent
superluminal velocities.
\section{Astrometric model for point-like sources} \label{sec:astrometric}
The aim of this section is to provide expressions that complete the
astrometric models at $\mu as$ accuracy incorporating the LTT. Under
the assumptions of Section \ref{sec:trajectories} and Section
\ref{sec:delay}, the observed direction of a point-like source given
by an observer at rest at the event of observation is
\begin{eqnarray}\label{eq:direction}
  \rvec\left[\tobs,\te \right] &=& \left< \xlrm + \vlrm \Dte \right. \\
   &+& \left.\D\left[\teO + \Dte\right] - \xobs\left[\tobs\right] \right>
   \nonumber
\end{eqnarray}
\noindent The next subsections provide general purpose
parametric expressions to be used in the astrometric model for a
point-like source outside the solar system. We applied the
formalism of the \emph{local triad} as described in
\citet{murray:1983}. Section \ref{sec:NLM} extends the astrometric
model to sources in nonlinear motion including the LTT amplitudes.
At this point, the expression of the observed direction
(\ref{eq:direction}) depends on the emission instant $\te$.We will
include the LTT in the astrometric model using just the relation
between $\Dte$ and $\Dtobs$ given by (\ref{eq:simple}) into
(\ref{eq:direction}).
\subsection{Linear Reference Motion}
\label{sec:LRM}
This is the simple case where the source is moving in linear motion
(i.e. a single star). Substituting (\ref{eq:simple}) in
(\ref{eq:direction}) and imposing $\D\left[t\right] = 0 \, \forall\
t\ $, we obtain
\begin{eqnarray} \label{eq:astroLRM}
\rvec_{LRM}\left[\tobs\right] &=&  \left< \lO\left(1 + \mu_{r0}  \Delta T_{LRM}\right) \right.\\
                         &+& \left( \mu_{\delta0} \pvec +
                         \mu_{\alpha0}^{*} \qvec
                      \right) \Delta T_{LRM} \nonumber \\
              &+& \left. \pivec \left[\tobs\right] \right> \nonumber \, ,
\end{eqnarray}
\begin{eqnarray}
\pvec &=& \left( -\sin \delta_0\, \cos \alpha_0,\,\, \right.\\
      & &        -\sin \delta_0\, \sin \alpha_0,\,\, \nonumber \\
      & & \left.  \cos \delta_0
          \right) ,
          \nonumber \\
\qvec &=& \left( \sin \alpha_0,\,\, \right.\\
      & &        \cos \alpha_0 ,\,\, \nonumber \\
      & &\left. 0
         \right) ,
         \nonumber \\
\lO   &=& \left( \cos \alpha_0\, \sin \delta_0,\,\, \right.\\
      & &        \cos \alpha_0\, \sin \delta_0,\,\, \nonumber \\
      & &\left. \sin \delta_0
         \right) ,
         \nonumber \\
\mu_{\alpha0}^{*} &=& \mu_{\alpha0} \ \cos \delta_0 =\frac{\a\vlrm \cdot \qvec}{\Xlrm}\, ,\\
\mu_{\delta0}     &=& \frac{\a\vlrm \cdot \pvec}{\Xlrm}\, , \\
\mu_{r0}          &=& \frac{\a\vlrm \cdot   \lO}{\Xlrm}\, ,  \\
\pivec\left[\tobs\right]   &=& \frac{\xobs\left[\tobs\right]}{\Xlrm} = \frac{\Pi_0}{AU}\xobs\left[\tobs\right]\, ,
\\
\Delta T_{LRM}   &=& \Dtobs +
                  \frac{1}{c} \ \lO \cdot \Dxobs \,   \label{eq:tlrm}.
\end{eqnarray}
\noindent These definitions extend those given for the HIPPARCOS
catalog \citep[see][vol.~1]{HIP} and include the Roemer correction
due to observer's motion, which was already introduced by
\citet{kopeikin:1992}. The vectors $\pvec$ and $\qvec$ are unit
vectors tangent to the celestial sphere at $\lO$ direction, pointing
towards the direction of increasing \emph{declination} and
\emph{right ascension} respectively. The quantities $\alpha_0$,
$\delta_0$, $\Pi_0$, $\mu_{\alpha 0}^{*}$, $\mu_{\delta 0}$ and
$\mu_{r0}$, are the so-called \textit{Barycentric astrometric
parameters} for a point-like source in rectilinear motion at the
barycentric reference epoch $\tobsO$. The angles $\alpha_0$ and
$\delta_0$ are the right ascension and the declination of the
equatorial coordinate system given in $rad$. The parameter $\Pi_0$
is the parallax in radians. The symbol $\mu_{\alpha 0}^{*}$ is the
proper motion in the $\alpha_0$ direction multiplied by $\cos
\delta_0$, which gives the correct angular shift correcting the
distortion of the spherical coordinates towards the poles, and
$\mu_{\delta 0}$ is the proper motion in the declination direction;
both expressed in $rad\, s^{-1}$. The parameter $\mu_{r0}$ is
known as \textit{astrometric radial velocity}
\citep[see][]{lindegren:2003} given in $s^{-1}$. $AU$ is the
Astronomical Unit which is currently defined as a constant.
\subsection{Nonlinear Motion} \label{sec:NLM}
The expression that generalizes to point-like sources in nonlinear
motion is straightforward. The six astrometric parameters described
in Section \ref{sec:LRM} are used to define a fiducial LRM while
$\D$ contains the nonlinear contributions,
\begin{eqnarray} \label{eq:astroNLM}
\rvec\left[\tobs\right] &=&  \left< \lO\left(1 + \mu_r \Delta T \right)  \right. \\
                        &+&  \left( \mu_{\delta0}     \pvec +
                   \mu_{\alpha0}^{*} \qvec
                \right) \Delta T \nonumber  \\
                &+& \frac{\D\left[\teO + \a\Delta T\right]}{\Xlrm} +
                            \left.\pivec\left[\tobs\right]  \right> ,\nonumber \\
         \nonumber \\
\Delta T &=& \Dtobs \label{eq:t}\\ 
      &-& \frac{1}{c} \ \lO \cdot \left(
        \DDteto - \Dxobs \right)
      \, \, \nonumber .
\end{eqnarray}
\noindent These expressions are sufficient to include LTT in an
astrometric data reduction algorithm at $\muas$. Classically, the
LTT terms with $\D$ inside $\Delta T$ were \textit{safely} neglected
since the astrometric measurements were not precise enough. In the
next section we will analyze the effect of this LTT term on the
observed direction, which is the LTT astrometric signature.
\section{Analytic estimation of the LTT signature}
\label{sec:analytic}

To estimate the astrometric LTT signature we need to compare
(\ref{eq:astroNLM}) with the \emph{classical} approach for the
observed direction $\rvec_c$. As the \emph{classical} approach we
define

\begin{eqnarray} \label{eq:classicNLM}
\rvec_c &=&  \left< \lO\left(1 + \mu_r \Delta T_c \right) \right. \\
                &+&
                            \left( \mu_{\delta0}     \pvec +
                   \mu_{\alpha0}^{*} \qvec
                \right) \Delta T_c \nonumber \\
                &+&
                \frac{\D\left[\teO + \a\Delta T_c\right]}{\Xlrm} +
                            \left. \pivec\left[\tobs\right]  \right> \, ,
         \nonumber \\
\Delta T_c &=& \Dtobs \, ;
\end{eqnarray}

\noindent where the difference with respect to (\ref{eq:astroNLM})
is essentially in $\Delta T_c$, which \textit{classically} does not
include the LTT contribution due to the nonlinear motion of the
source with respect to the LRM, and due to the position of the
observer with respect to the barycenter of the solar system. This
last contribution (R\"{o}mer term) is already contained in recent
accurate astrometric models \citep{kopeikin:1992}, but we prefer to
also consider it here for completeness. The comparison of $\rvec$ and
$\rvec_c$ is performed by direct subtraction of
(\ref{eq:astroNLM}) and (\ref{eq:classicNLM}) up to $\OO2$,
considering $\OO1$ all those terms containing expressions
proportional to those in
(\ref{eq:smallintervals})--(\ref{eq:small}). Then, the astrometric
LTT signature $\delta\rvec$ is defined as

\begin{eqnarray}
\delta \rvec\ &=& \rvec\left[\tobs\right] - \rvec_c\left[\tobs\right] \, ,
\end{eqnarray}

\noindent and, after some algebra

\begin{equation}
\delta \rvec = \lO \X \left( \ve{\delta}\X \lO \right) + \OO3\  \label{eq:LTTfinal1} \\
\end{equation}
\begin{equation}
\ve{\delta}  = - \a\ \frac{\vlrm + \V\left[\teO +
\a\Dtobs\right]}{\Xlrm} \Delta T_{LTT} \, \label{eq:LTTfinal2}
\end{equation}
\begin{equation}
\Delta T_{LTT} = \frac{\lO \cdot \left( \DDteto - \Dxobs \right)}{c}
\, .
\end{equation}

\noindent The presence of the \textit{orbital velocity} vector $\V$
is justified in Appendix \ref{sec:algebra1}. It is called orbital
velocity by direct analogy with the binary system. The LTT shift
$\delta\rvec$ is an astrometric detectable quantity since it is in
the perpendicular direction to the line of sight $\lO$, as is
explicit in (\ref{eq:LTTfinal1}). The LTT shift depends on the
radial projection of the \textit{orbital} motion $\lO \cdot \DD$.
This implies that using accurate astrometric measurements one would, in
principle, be able to constrain the radial motion of the
source or, on the other hand, if some information of the radial motion of
the source is provided (i.e. radial velocities), the fit of the
astrometric orbit might be more robust and accurate. In Section
\ref{sec:real}, the relevancy of the astrometric LTT signature will
be shown in some well known multiple systems. The \emph{apparent}
position of a source is advanced or delayed with respect to the
\textit{nonretarded} trajectory. That is why the expression found in
(\ref{eq:LTTfinal2}) is just the \emph{instantaneous proper motion}
multiplied by the time interval $\Delta T_{LTT}$, which is time
dependent.

In formula (\ref{eq:LTTfinal2}) two velocities appear instead of the
instantaneous total velocity. They are kept separate since the ways
they affect the observed direction (their \emph{astrometric
signature}) have quite different properties. The \emph{proper
motion} term $\vlrm$ is constant and the time dependence of this
part of the astrometric LTT signature will come only from the
variations of the $\Delta T_{LTT}$ interval. However the
\emph{orbital velocity} $\V$ is intrinsically time dependent,
periodic in most cases, and its coupling with the $\Delta T_{LTT}$
interval will produce a more \emph{sophisticated} astrometric
signature. As shown in Section \ref{sec:real}, when applied to
binary systems, the scaling law of each contribution with respect to
the orbital elements is significantly different.

An additional comment. The term $\D\left[\teO\right]$ may be included in
$\xlrm$ by imposing $\D\left[\teO\right]=0$. However, it happens that 
this assumption may not be useful in practical cases (i.e. when applied 
to fitting the orbital parameters of a binary system). We will keep the 
current expression unless we need to implement LTT in the particular
modelling of an object.

The LTT signature is one among other astrometric effects becoming relevant at
second order astrometric accuracy. These effects include perspective
acceleration (or astrometric radial velocity), and all kind of couplings of the
parallax with the proper motion or the non-linear motion. A comprehensive list 
can be found in \citet{dravins:1999}. All of them are 
naturally included in (\ref{eq:astroLRM}) and (\ref{eq:astroNLM}), since our
expressions are directly derived from the kinematical model of 
the source. Expanding (\ref{eq:astroNLM}) up to $\OO2$ in the small terms
(\ref{eq:small}), the vectorial expressions of all such contributions are
explicitly obtained. Any astrometric study aiming to obtain information using
any sort of second order contributions must properly consider the LTT signature
explained in this work.

\section{Some numerical estimates} \label{sec:real}

Let us naively use the expression (\ref{eq:LTTfinal2}) to obtain
some order of magnitude estimates of the LTT signature. The
semiamplitudes (denoted by $\overline{\delta\rvec}$) of the
astrometric LTT signatures for a component in a binary system in
circular orbit can be estimated as

\begin{eqnarray}
\overline{\ve{\delta}\rvec}_{proper}  \sim \ 15.812 &\muas&
\frac{a\arcsec \mu_{mas/year}}{\pi_{mas}} \ \sin i \label{estimation1}\ , \\
\overline{\ve{\delta}\rvec}_{orbital} \sim 99\,353 &\muas&
\frac{a\arcsec^{2}}{\pi_{mas}P_{year} } \ \sin i   \label{estimation2}.
\end{eqnarray}

\noindent Just for simplicity, these expressions are obtained imposing 
$\Delta\xobs = 0$. Relations (\ref{estimation1})--(\ref{estimation2}) are 
provided using catalog-like parameters where $a\arcsec$ is the projected 
semi-major axis in $arcseconds$, $\mu_{mas/year}$ is the proper motion module in
$mas\, year^{-1}$ and $\pi_{mas}$ is the parallax in $mas$. The effect is
modulated by the inclination of the orbit.

Kepler's third law can be used to obtain the expressions
(\ref{estimation1}) and (\ref{estimation2}) as powers of the orbital period 
$P$ and the orbital semimajor axis $R$. It is found that
$\overline{\ve{\delta}\rvec}_{proper}$ scales as $R^{1}$ (or $P^{2/3}$), while
$\overline{\ve{\delta}\rvec}_{orbital}$ scales as $R^{1/2}$ (or $\ P^{1/3}$). This illustrates that for systems with long period orbits
the LTT signature due to the coupling with the \emph{proper motion}
will be more significant. This is expected since $\vlrm$ does not
depend on the semimajor axis of the orbit(it is only related to
the velocity of the center of mass of the system), while $\V$
becomes smaller at larger orbital distances (i.e. in the solar system,
distant planets have lower orbital velocities).

The semiamplitudes obtained using (\ref{estimation1}) and
(\ref{estimation2}) for some nearby systems are provided in
Table~\ref{tab:num}.

\begin {table*} [htb]
\center
\begin{tabular}{lccccccc}
\hline\hline

             System
           & $Period$
           & $a''$
           & $i$
       & $Parallax$
       & $Proper\ motion$
       & $\Delta \s_{proper}$
           & $\Delta \s_{orbital}$
       \\

       \
           & $years$
           & $arcsec$
           & $deg$
       & $mas$
       & $mas/year$
           & $\mu as$
       & $\mu as$

\\
\hline
61 Cyg
& $722.0$
& $14.9$
& $51.85$
& $294$
& $5227$
& $3293$
& $81.7$ \\
$\alpha$ Cen
& $79.9$
& $8.75$
& $79$
& $742$
& $3672$
& $670$
& $415$ \\
HD 110314
& $3.09$
& $0.021$
& $122.9$
& $14$
& $191.9$
& $3.80$
& $0.86$ \\
HD 2475
& $5.65$
& $0.146$
& $64$
& $118$
& $31.01$
& $0.54$
& $2.85$ \\
AB pic-b
& $\sim 3000$
& $0.753$
& $??$
& $21.97$
& $47.36$
& $\sim 50$
& $\sim 1$ \\

\end{tabular}
\caption{The data in this table were prepared using the \textit{WDS
Sixth Catalog of Orbits of Visual Binary Stars}
\citep{catalog:sixth} and the HIPPARCOS catalog \citep{HIP}. The
value of some of the orbital parameters shown are average values.
The numbers shown here are orientative since the circular model
applied is very unrealistic. It can be seen that both contributions,
$\delta\rvec_{proper}$ and $\delta\rvec_{orbital}$,  may have quite
different signification depending on the binary system. In very
eccentric systems, like $\alpha$ Cen, the \emph{orbital} term can be
considerably larger than the numbers shown here around the
perihelion. Another feature that is blurred by the applied
approximations is that if the components of the system have
different masses, each component might show very different
astrometric LTT signatures. The values for AB pic-b are \emph{very}
approximated and are based on very recent and sparse data of this
candidate to planetary system \citep{chauvin:2005}. This values are added here
as an example of the applicability of the astrometric LTT signature
to general purpose orbital solutions. In this case, the LTT
signature would not be very useful to improve the orbital solution
due to the very long period and tininess of the signal.
}\label{tab:num}
\end{table*}

\begin{figure*}[htb]
\centering
\includegraphics[width=15.0cm]{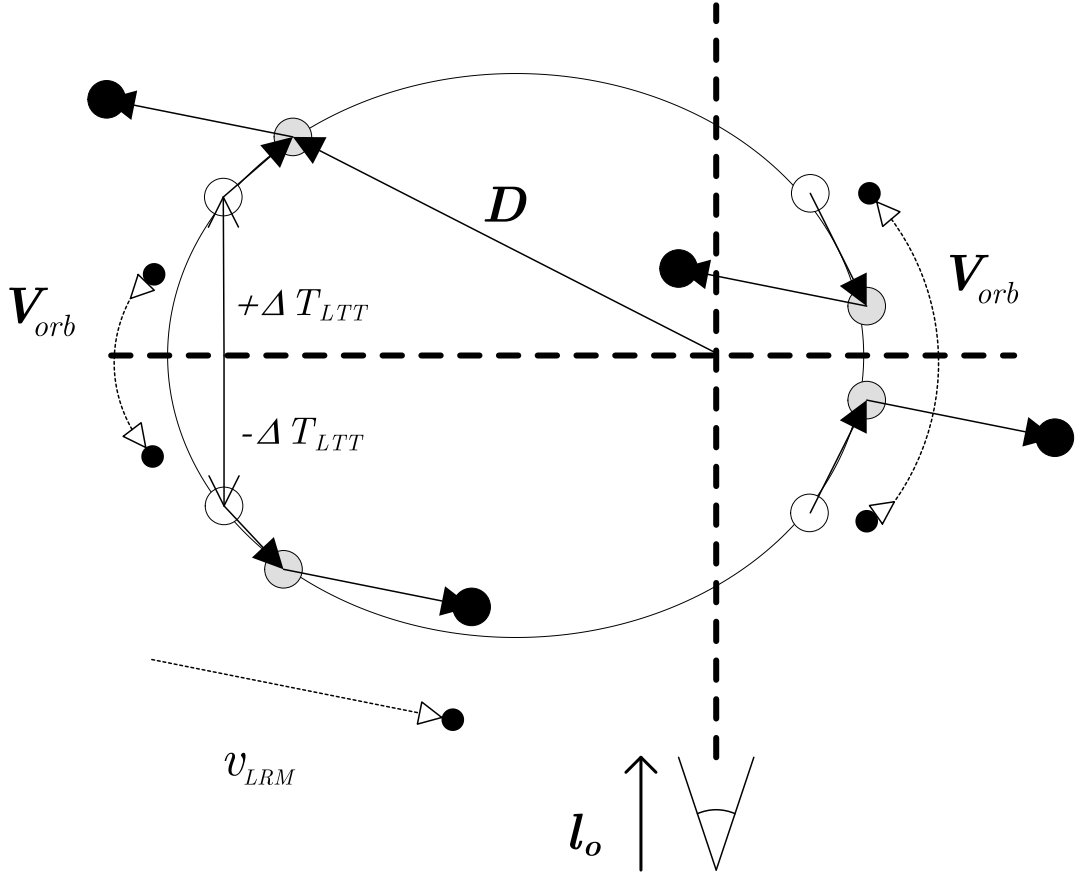}
\caption{Astrometric LTT signature in a binary system.  In the picture, 
the elliptic orbital motion of a companion is shown in the case where the
inclination is $90\deg$. The horizontal dashed line is the semi-major axis
of the orbit and is perpendicular to the line of sight given by $\lO$. The white
circles represent the nonretarded positions of the star at different moments
of the orbit. The gray circles are the apparent positions corrected by the
Light-Travel Time appliying only the LTT correction due to the instantaneous
orbital velocity $\V$. The black circles are the final apparent position after
also considering the motion of the barycenter of the binary system.
The proper motion of the system is omitted to simplify the
visualization of both the LTT signatures. The pure orbital correction $\V
\Delta T_{LTT}$ changes its orientation continuously depending on
the orbital position(gray circles) and it is larger when the object
is closer to the center of mass of the system (right side of the
figure). On the other hand, the proper motion contribution $\vlrm
\Delta T_{LTT}$ (dark circles) always contributes in the same
direction changing only the sign and the amplitude of the perturbation. The
projection of the shifts on the plane perpendicular to the line of sight
are the astrometrically measurable quantities.
}
\label{figDelay2}
\end{figure*}

A detailed model to include the astrometric LTT signature in a
strict orbital solution of binary systems will be provided in a
later work which is in preparation, since some detailed discussion
of the definition of the orbital parameters must be provided in order to keep
the astrometric model accuracy at the $\mu as$ level. We prefer to omit this
discussion here for the sake of brevity. 

As shown in Table~\ref{tab:num}, long period binaries habitually have 
\textit{larger} LTT signatures coming from the coupling with
the \emph{proper motion} term. This is the case of $61$ Cyg. In that
situation, the LTT signature could be resolved using available
long-term lower precision astrometry. 

\section{Conclusions} \label{sec:conclusions}

We have shown that the LTT must be taken into account in order to
obtain accurate models of the observable quantities in precise
modern astrometry. Furthermore, since LTT is an object dependent
effect, precise relative astrometric measurements are sufficient to
resolve the astrometric LTT signatures. The astrometric LTT signature
may be used in highly precise ground based observations to obtain
additional information on a given object. The expressions derived in
(\ref{eq:LTTfinal1})--(\ref{eq:LTTfinal2}) can be applied using
$O-C$ techniques with the available astrometric data. Furthermore,
it is clear that LTT must be taken into account in the
interpretation of the data obtained by the planned space astrometric
missions which will attempts to reach astrometric accuracies of a few $\mu
as$. As an example, the coupling of the R\"{o}mer delay with the
proper motion and the orbital velocity of a star can mimic the
astrometric wobble caused by a planetary mass object with a period
of around one year (considering an observer traveling on the
vicinity of the earth) if LTT is not properly included in the
astrometric model.

It has been found that the LTT signature is boosted by the proper
motion of the system. Multiple systems with high proper motions
(thick disk, globular clusters, nearby halo objects) might be
objects of investigation if properly adapted astrometric models are used.

Another conclusion of interest is that if the astrometric LTT
signature is large, information on the radial geometry of a system
can be obtained. In the case of the binary systems this information allows us
to solve (or at least constrain), the full set of orbital elements
without information about radial velocities. Despite of that, the
numbers appearing in Table~\ref{tab:num} are not very impressive
(even at $\mu as$ accuracy level). It seems unfeasible to use the
astrometric LTT signature to improve significantly the knowledge of
a given system since radial velocity curves are much better
measured. The suggestion here is to use the radial velocity
measurements as an input to improve the astrometry of a given source
using the LTT corrected orbital description. How to proceed properly 
in such cases is under investigation.

When imaging capabilities of exoplanetary systems become available,
the fitting of an LTT orbit might lead to constraining the size of the
orbit in the line-of-sight direction without the use of radial
velocity measurements, which may be impracticable on such faint
objects unless the interferometric techniques improve
significantly.

The relations (\ref{estimation1})--(\ref{estimation2}) can be used
as a good indicator of whether LTT astrometric effects must be taken
into account for more general objects (open clusters, globular
clusters, galaxies, fast orbiters around massive black holes) in
nonlinear motion. For this purpose we define the \textit{LTT
astrometric signal} as

\begin{eqnarray} \label{adimensional}
 \mathcal{L}_{TT} \equiv {V\over{c}} {L\over{d}},
\end{eqnarray}

\noindent where $V$ and $L$ are the characteristic \textit{velocity}
and \textit{longitude} of the system, respectively; $c$ is the speed
of light in the same units as $V$; and $d$ is an estimation of the
distance to the source given in the same units as $L$.
$\mathcal{L}_{TT}$ is an adimensional quantity that can be directly
interpreted as an angle(in radians). If this number is of the order
of the astrometric accuracy used to describe an astronomical object
outside the solar system, then the LTT should be taken into account
in order to give a correct interpretation of the astrometric data.

For some purposes (such as the construction of an astrometric catalog) it
is sometimes useful to make the abstraction that at the reference
epoch $\tobsO$ the observer is located at the Barycenter of the
Solar System. If this is done, the initial direction
$\rvec\left[\tobsO\right]$ will not depend on the R\"{o}mer delay
$\lO \cdot \xobs\left[\tobsO\right]$ due to the observer position at
$\tobsO$ as it does in expressions (\ref{eq:t})--(\ref{eq:tlrm})
or in the astrometric LTT signature formula (\ref{eq:LTTfinal2}).
Then we can substitute
$\xobs\left[\tobs\right]-\xobs\left[\tobsO\right]$  by simply
$\xobs\left[\tobs\right]$ in both equations (\ref{eq:tlrm}) and
(\ref{eq:t}) or in (\ref{eq:LTTfinal2}). This consideration is very
useful if you want to create an astrometric catalog for a given
reference epoch absolutely independent of the relative initial
position of observer $\xobs\left[\tobsO\right]$ and the initial
direction of each source $\lO$.

A precise modelling of the LTT to be applied in the astrometry of
binary systems is being developed and will be presented in a later
publication.

\nocite{DARWIN:2000}
\nocite{kaplan:2005}

\begin{acknowledgements}
This work was carried out under the financial support of
grant ESP2003-04352  of Ministry of Education and Science, Spain.
The authors thank I.Ribas and S.A.Klioner for motivating the development of
this work with their comments. All the members of the GAIA group in
Barcelona are gratefully acknowledged fow allowing us to develop this
work in close contact with the Gaia scientific community.
\end{acknowledgements}

\bibliographystyle{aa}
\bibliography{Paper}

\appendix

\section{First order derivation of Equation of time delay}
\label{sec:algebra1}

The first term on the right side of the equation
(\ref{eq:delay}), to first order as defined in 
(\ref{eq:smallintervals})--(\ref{eq:small}), reads

\begin{eqnarray}
& &\frac{\left| \xobs\left[\tobs\right]-\xe\left[\te\right] \right|}{c} \ap \frac{1}{c}\ \Xlrm  \label{devel2} \\
& & \,\,\,\,\, + \,\, \lO \cdot { \vlrm \Dte \over{c}} \,\, + \,\, \lO \cdot \frac{\D \left[\te\right]}{c}  \nonumber \\
& & \,\,\,\,\, - \,\, \lO \cdot \frac{\xobs\left[\tobs\right]}{c} \,\, + \,\, \OO2 \nonumber ;
\end{eqnarray}

The second term of the right hand of (\ref{eq:delay}) is
straightforward using $\teO$ instead of $\te$ in the last expression
(\ref{devel2}). After some algebra the emision time interval $\Dte$ can be
written as

\begin{eqnarray} \label{almost}
\Dte &=& \frac{1}{1+ \lO \cdot \frac{\vlrm}{c}}
\left(\, \,
\Dtobs \right. \\
&-&        \lO \cdot \frac{\DDte}{c} \nonumber \\
&+& \left. \lO \cdot \frac{\xobs\left[\tobs\right] - \xobs\left[\tobs^{0} \right]}{c}\     \ \right) \nonumber \\
&+& \OO2 \nonumber ;
\end{eqnarray}

\noindent The term multiplying the full expression is responsible for apparent
superluminal velocities. For this reason we call this term
superluminal factor
\begin{eqnarray} \label{eq:superlumdemo}
\a &=& \frac{1}{1+\lO \cdot {\vlrm \over{c}}} ;
\end{eqnarray}

In spite of the suppression of the second order terms there is still a
dependency on $\te$ on the right hand of equation(\ref{almost}) in $\D$.
This equation is enough to solve $\Dte$ iteratively. But our purpose
is to obtain a closed form accurate to $\OO1$. To solve this we
consider

\begin{eqnarray}
\DDte &=& \D \left[
               \te^{0} + \a \Dtobs + \e\right] \\
       &-& \D\left[\te^{0}\right], \nonumber
\end{eqnarray}
\begin{eqnarray}
\e = &-&\a\,\,\lO \cdot \frac{\DDte}{c}\   \\
     &+&\a\,\,\lO \cdot \frac{\xobs\left[\tobs\right]-\xobsO\left[\tobsO\right]}{c}\
              \sim \OO1 \nonumber;
\end{eqnarray}

Taking this into account, we can write to first order in $\e$

\begin{eqnarray}
& &\frac{\DDte}{c} = \label{eq:vdevel} \\
& &\, \, \,\,\,\,\,\, \frac{\D\left[\teO + \a \Dtobs\right] - \D\left[\teO\right]}{c} \nonumber \\
& &\, \, \, + \frac{\V\left[\teO+\a\Dtobs\right]}{c}\,\, \e + \OO2 \nonumber
\end{eqnarray}

\noindent In (\ref{eq:vdevel}) the term $\frac{\V}{c}\e$ is of
$\OO2$ and will be neglected in (\ref{almost}). The
development (\ref{eq:vdevel}) is also used to justify the appearance
of $\V$ in equation (\ref{eq:LTTfinal2}). These are all the
ingredients needed to obtain the relation of $\Dte$ in terms of
$\Dtobs$ to first order

\begin{eqnarray} \label{LTT1}
\Dte &=& \a \left( \Dtobs \right. \\
         &-&
         \lO \cdot \frac{\D\left[\teO + \a \Dtobs\right] - \D\left[\teO\right]}{c}\   \nonumber \\
         &+& \left.
         \lO \cdot \frac{\xobs \left[\tobs\right])-\xobs \left[\tobsO\right]}{c}\
     \right) \nonumber \\
     &+& \OO2  \nonumber
\end{eqnarray}

\noindent which is the same as equation (\ref{eq:simple}) with the
suitable notation shortcuts explained in
(\ref{eq:superlum})--(\ref{eq:DDteto}).

\end{document}